

\magnification=1200
\baselineskip=20pt
\ \
\def\cl{\centerline}

\hyphenation{Schwarz-schild}

\vskip 5.0 cm
\cl{\bf MAGNIFICATION BIAS IN GALACTIC MICROLENSING SEARCHES }
\vskip 3 cm
\baselineskip=15pt
\cl { Robert J. Nemiroff }
\bigskip
\bigskip\bigskip
\cl{ George Mason University, CSI, Fairfax, VA 22030 }
\cl{ NASA / Goddard Space Flight Center, Code 668.1, Greenbelt, MD
20771 }

\bigskip\bigskip\bigskip

\cl{ In Press: }

\cl{ The Astrophysical Journal }

\vfill\eject

\baselineskip=20pt
\parindent=20pt
\parskip=5pt

\cl{\bf ABSTRACT }

It is shown that a significant amount of detectable gravitational
microlensing events that could potentially be found by MAssively Parallel
Photometry (MAPP) projects (such as the MACHO, EROS and OGLE
collaborations) will occur for stars too dim to be easily noticed
individually by these projects. This is the result of a large magnification
bias effect, a bias of including high magnification events in any
flux-limited sample. The probability of detecting these events may be as
high as 2.3 times the lensing probability of stars currently being
monitored by MAPP collaborations.

\noindent
{\it Subject Headings:} stars: low-mass, brown dwarfs, Galaxy: halo, dark
matter, gravitational lensing

\vfill\eject

\cl{\bf 1. Introduction }

Recently Alcock et al. (1993, the MACHO collaboration), Aubourg et al.
(1993a, the Experience de Recherche d'Objets Sombres or EROS project) and
Udalski et al. (1993a, called Optical Gravitational Lensing Experiment or
OGLE) have all reported seeing light curves of stars indicative of a
fainter star passing in front of - and hence gravitationally magnifying -
the light from a background star: a gravitational microlens event.  The
probability of seeing such an event was predicted originally by Paczynski
(1986), Griest (1991), and Nemiroff (1991) based on the probability
formalism developed in Nemiroff (1989).  Assuming that the nature of these
events is correctly identified, these events are measuring the mass and
density of stars in our disk and dark-matter halo objects in the Galactic
halo.

MAssively Parallel Photometry (hereafter referred to as MAPP) projects
currently monitor known stars for variations indicative of gravitational
lensing.  The method currently used by the MACHO collaboration, as
described in Alcock et al. (1992), the EROS collaboration, as described in
Aubourg et al. (1993b), and the OGLE project, as described in Udalski et
al. (1993b), appears to disregard lensing of stars that are too dim to be
easily noticed in the absence of a magnification event.

The general luminosity function of stars slightly dimmer than those being
monitored by current MAPP projects clearly shows that there are
increasingly more stars at dimmer magnitudes.  Although more numerous,
these dim stars must undergo a stronger lens effect to be even potentially
noticeable by a MAPP search. The first group to consider the effects of
these dim stars was Bouquet, Kaplan, Melchior, Giraud-Heraud, and Baillon
(1994).  They presented numerical Monte-Carlo results showing that
monitoring pixels instead of stars could provide a ten-fold increase in
detection efficiency, particularly in the EROS CCD program.  It will be
shown here by simple analytical estimates that the probability of detection
of lensing of these dim stars is indeed substantial for all of the MAPP
projects, although the probability is a strong function of the luminosity
function of the source stars involved, which has significant uncertainty.

A possible future MAPP mission of M31 has been suggested by Crotts (1992).
Crotts' suggested procedure for finding M31 microlensing is similar to the
proposed method here in that it depends on the monitoring of a whole
stellar-rich field rather than monitoring a previously determined list of
stellar lensing candidates.  In this work we apply this ``field monitoring"
concept to the present MAPP missions monitoring the LMC and the galactic
bulge. We show explicitly that the lens detection rate is a function of the
monitoring rate and the luminosity function of stars slightly below the
detection threshold in the source field. This this type of lensing
phenomenon is well known in gravitational lensing as ``magnification bias."
We estimate the potential increase in lens detections of the currently
running MAPP missions.

Section 2 will discuss generally the probability of these dim stars being
detectably gravitationally lensed.  Section 3 will apply this analysis to
currently operating MAPP projects.  Section 4 will provide discussion.

\bigskip

\cl{\bf 2. Magnification Bias and MAPP Searches }

Magnification bias occurs for a flux-limited sample by the natural
inclusion of sources normally too faint to make the sample but included
fortuitously because of being magnified by a foreground gravitational lens
(Blandford and Narayan 1992; Narayan 1994). This bias is particularly
important for sources with a luminosity function which shows a great
increase in the number of sources marginally fainter than the flux-limit.

Nemiroff (1991) showed that for sources monitored on a time-scale short
compared to the duration of a lens event, the waiting time between lensing
events of magnification factor $A$ is, when $A$ is large,
 $$ t_{wait} \propto {A \over N} ,
 \eqno(1)$$
where $N$ is the number of stars being monitored. This waiting time depends
on the relative motion of the lens and the sources and therefore its
dependence on $A$ is qualitatively different than discussed in the static
cases of Paczynski (1986), Griest (1991) and Nemiroff (1991), although they
note that the classically defined optical depth to gravitational lensing is
small allowing single lens models.  Note also that although the probability
of lensing above a given magnification is proportional to optical depth,
optical depth itself does not include relative lens and source motion and
is tied to a specific magnification.  Therefore optical depth will not be
explicitly considered here.  Because of this, I will digress and discuss a
simple method to see Eq. 1 intuitively.

Consider a simple adaptation of classical mean free path arguments to
gravitational lensing probabilities.  Around each lens, picture an angular
circle within which a background star must pass to undergo a detectable
magnification.  Now consider the relative angular motion between lenses and
background stars as potentially visible to the observer. A background star
will experience this relative angular motion and therefore has a two
dimensional ``mean free path" $l$ before encountering a circle around a
lens which will magnify it detectably. In general, for a single source, $l
\propto (\sigma B)^{-1}$ where $\sigma$ is the (unknown) angular
two-dimensional number density of lenses, each with `lens detection'
cross-section $B$. For large magnification events, $B \propto A^{-1}$, so
that $l \propto A$. For a canonical relative angular velocity between
lenses and sources, the mean time between events $t_{wait} \propto l$ so
that $t_{wait} \propto A$. Clearly the waiting time between lensing events
for $N$ sources scales as $1/N$, so that the form or Eq. (1) is verified.
The probability that a source in an ensemble of $N$ sources will undergo
detectable lensing is directly proportional to the size of the ensemble and
inversely proportional to the waiting time, so that $P \propto N / A$.

We now assume that there is a strict limiting magnitude for source stars
above which a MAPP search will surely monitor a star and below which a MAPP
search will definitely not monitor a star.  To simplify terminology, we
will adopt the term ``bright" to refer to stars brighter this limiting
magnitude, and the term ``dim" to refer to stars dimmer than this
magnitude.

For a dim star to be detectably lensed, we will assume that the star must
be magnified by a factor $A_{dim}$ to reach the magnitude limit that
divides the bright and dim groups, and an additional magnification factor
$A_{MAPP}$ for lensing to be strong enough to be detectable.  One might
immediately think that lensing of the dim group is quite unlikely, relative
to the bright group, but this may be offset by the greater number of stars
in the dim group.

{}From the above assumptions, a dim star originally a factor $A_{dim}$ below
a MAPP detection limit could potentially be seen to undergo detectable
lens magnification if it is magnified by at least the amount $(A_{MAPP} +
A_{dim})$.  For stars well below the detection limit $A_{dim}$ will be
considered much greater than $A_{MAPP}$, so that $P_{dim} \propto N_{dim} /
A_{dim}$.  Now $A_{dim}$ is equivalent to the brightness of the detection
limit divided by the brightness of the dim star: $L_{lim}/L_{dim}$, where
$L_{lim}$ is the limiting absolute luminosity of the MAPP region.
Assuming that the luminosity function can be characterized as $N \propto
L^{\alpha}$ so that $N_{dim}/N_{lim} = (L_{dim}/L_{lim})^{\alpha}$ then
 $$ { P_{dim} \over P_{lim}  }
            \propto { (N_{dim}/N_{lim}) \over A_{dim} }
            \propto  L_{dim}^{\alpha + 1} .
 \eqno(2)$$
If the source stellar luminosity function has $\alpha < -1$, the
probability of dim star lensing actually {\it increases} as the apparent
brightness of the stars decrease.  This shows that magnification bias is
not only important - it may help create a more likely form of lens
detection.

If Eq. (2) continued to even dimmer stars indefinitely, dim star lensing
would be very much {\it more} probable than detection of lensing from the
bright stars.  In this case, the dimmer the star, the more of them there
are, the higher the magnification needed for detection, but the more
detections per time there should be.  Practically, this should continue
until the luminosity function of stars in the MAPP project becomes flatter
than $L^{-1}$, at which point the decrease in the probability of high
magnification microlensing becomes more important than the increase in the
number of stars at fainter magnitudes.

\bigskip

\cl{\bf 3. Application to Existing MAPP Projects }

Alcock et al. (1992) give a limiting magnitude for the MACHO collaborations
photometry of about 19.5 for a 5 minute exposure (at full moon). Aubourg et
al. (1993b) estimate a mean visual magnitude for their photographic search
of 19.0.  Udalski et al. (1993b) estimate a limiting visual magnitude for
the OGLE project of about 19.5. We assume a distance modulus in the $V$
band to the LMC of $V = 18.5$ (Blaha and Humphreys 1989). Therefore the
absolute visual magnitude of an LMC star at the limiting magnitude of the
MACHO and EROS collaborations is about $M_V = 1$.

Unfortunately, the intrinsic luminosity function of stars of this apparent
magnitude and fainter has significant uncertainties.  We first consider the
luminosity function of the LMC stars.  Table 4-5 in Mihalas and Binney
(1981), drawn from results obtained by Luyten (1968), gives the general
luminosity function of stars in the Galactic disk.  We will follow Blaha
and Humphreys (1989), who showed that the LMC and the Galactic disk have
similar luminosity functions for O and B stars in the range $-9 < M_V <
-7$, and assume that this similarity applies to dimmer stars as well.

We use Eq. (2) to estimate the probability of dim star microlensing by the
luminosity function until $M_V = 10$,  which is well past the magnitude
where additional lensing probability is substantial.  This calculation
shows that there should be about 2.3 times the number of lens events (of
any magnification) involving stars too dim to be continually monitored than
involving stars currently monitored by the MACHO or EROS collaborations.

We note that the Luyten luminosity function's slope rises above -1
coincidently at $M_V = 1$, the magnitude limit of the MACHO collaboration
search. Were the MACHO and EROS searches not as deep, the magnitude of the
magnification bias would be expected to be much stronger. As it is, were
EROS's magnitude limit fully 1/2 magnitude brighter than MACHO's, EROS
would be susceptible to about an additional 25 \% of lensing events
occurring on stars down to only 1/2 magnitude below this limit.

Things do not look as promising for the OGLE project in this respect. The
luminosity function of stars in Baade's window, as measured by Holtzman et
al. (1993), again coincidently shows a break near the completeness limit,
but this time at a slightly dimmer magnitude of about $M_V = 20.0$. Stars
dimmer than this break have a luminosity function too flat for them to be
significant lens candidates.  Between the OGLE magnitude limit of 19.5 and
this break, however, there might be a significant magnification bias
effect.  From Figure 5 of Holtzman et al. (1993), we estimate a luminosity
function  with slope of about -1.25 in the range $19.5 < M_V < 20.0$,
although the plotted error bars show significant uncertainty in this slope.
{}From Eq. (2) we would estimate that there should be about an equal number
of detectable lens events (of any magnification) involving stars too dim to
be continually monitored than involving stars currently monitored by the
OGLE collaboration.

\bigskip

\cl{\bf 4. Discussion }

Lensing of stars below the detection threshold could be found in
anomalous brightening events not matched with known stars recorded by the
MACHO, EROS and OGLE collaborations.  The source stars involved with these
events might be recovered by a large telescope after the lensing event.
Current detection techniques necessarily select large magnification events
for these dim stars.   Since only during times of high magnification would
these events be visible, the duration of this type of lensing event would
be less than the ``normal" MACHO event. This reduced duration might result
in a potential ``duration" bias against the detection of these events.

It is interesting to note that the {\it static} probability of lensing
detection - that not assuming relative lens and source motion - would
predict that magnification bias is unimportant.  This is because
the static probability of lensing magnification is proportional to $A^{-2}$
(at large $A$), so that it is much less probable for high lens
magnification that the increase in the number of potential sources given by
the stellar luminosity function of the LMC would not compensate.  Only when
considering the {\it dynamic} probability - which includes relative motion
between the source and lens populations in the probability estimate, does
it become clear that magnification bias may create a sizeable effect.  For
the dynamic lens paradigm to be in effect, however, the time between
measurements of apparent stellar brightness must be small compared to the
duration of the lensing event.

Note that the probabilities estimated here assume that the source star had
to be magnified above the MAPP projects' current magnitude limit to be
considered a microlensing candidate.  Were a MAPP project begun using a
telescope with higher angular resolution, the magnitude limit would be
extended to dimmer stars, and each dim star need only undergo a
magnification of $A_{MAPP}$ to be considered a microlensing candidate.
Then the probability of detectable lensing would clearly rise much above
that reported here - it would then be just roughly proportional to the
increased number of stars being monitored.

Lensing of stars below a limiting magnitude works best when this magnitude
occurs at relatively bright intrinsic stellar luminosities.  At these
luminosities the number-magnitude relation is very steep - describing many
more stars in successively fainter magnitude intervals.  In this vein, a
proposed project for monitoring stars in M31 (Crotts 1994) would be quite
susceptible to this type of magnification bias effect.

Magnification bias comes about because of a magnitude limit, a magnitude
limit is usually used in MAPP searches because of source confusion, source
confusion results from the angular resolution limit not being able to
resolve stars from each other, and angular resolution limits at this level
result from the Earth's atmosphere.  Therefore were a space-based MAPP
project implemented, one might be able to extend the magnitude limit down
to a stellar brightness where the luminosity function would rise slow
enough such that magnification bias would not be important.  The main value
of a space-based MAPP project would be, of course, that many more stars
could be continually monitored for gravitational microlensing.  Failing
this, magnification bias will continue to be an important effect.

In sum, currently operating searches for gravitational microlensing by
stars in our Galaxy suffer a large magnification bias.  This bias
translates into a significant probability that stars too dim to be included
in an individual monitoring program would be magnified above detection
limits. Existing and future MAPP type searches might consider adapting
their data acquisition techniques and software to include the possibility
of detecting such events.

This work was supported by a grant from NASA.

\bigskip

\cl {\bf REFERENCES }
\parindent=0pt
\baselineskip=14pt
\parskip=6pt

Alcock, C. A. et al. 1992, Astron. Soc. Pac. Conf. Ser., 34, 193

Alcock, C. A. et al. 1993, Nature, 365, 621

Aubourg, E. et al. 1993a, Nature, 365, 623

Aubourg, E. et al. 1993b, Messenger, 72, 20

Blaha, C. and Humphreys, R. A. 1989, AJ 98, 1598

\hangindent=20pt
Blandford, R. and Narayan, R., 1992, Ann. Rev. Astron. Astrophys. 30, 311

\hangindent=20pt
Bouquet, A., Kaplan, J., Melchior, A.-L., Giraud-Heraud, Y., and Baillon,
P. 1994, in proceedings of ``The Dark Side of the Universe . . .", in press

Crotts, A. P. S. 1992, ApJ, 399, L43

\hangindent=20pt
Crotts, A. P. S. 1994, in Proceedings of the 31st International Liege
Colloquium: Gravitational Lensing in the Universe, in press.

Griest, K. 1991, ApJ, 366, 412

Holtzman, J. A. et al. 1993, AJ, 106, 1826

Luyten, W. J. 1968, MNRAS, 139, 221

\hangindent=20pt
Mihalas, D. and Binney, J. 1981, Galactic Astronomy, Structure and
Kinematics, 2nd Edition, (San Francisco, Freeman), Chapter 4

\hangindent=20pt
Narayan, R. 1994, in Proceedings of the 31st International Liege
Colloquium: Gravitational Lensing in the Universe, in press.

Nemiroff, R. J. 1989, ApJ, 341, 579

Nemiroff, R. J. 1991, A\&A, 247, 73

Paczynski, B. 1986, ApJ, 304, 1

\hangindent=20pt
Udalski, A., Szymanski, M., Kaluzny, J., Kubiak, M., Krzeminski, W., Mateo,
M., Preston, G. W., and Paczynski, B. 1993a, Acta Astron., 43, 289

\hangindent=20pt
Udalski, A. Szymanski, M., Kaluzny, J., Kubiak, M., and Mateo, M. 1993b,
Acta Astron., 43, 69

\vfill\eject
\end